\documentclass[3p,times,twocolumn]{elsarticle}

\usepackage{ecrc}

\volume{00}

\firstpage{1}

\journalname{Nuclear Physics B Proceedings Supplement}

\runauth{}

\jid{nuphbp}

\jnltitlelogo{Nuclear Physics B Proceedings Supplement}

\usepackage{amssymb}

\usepackage[figuresright]{rotating}
\usepackage{cancel}
\usepackage{amsmath}

\begin{document}

\begin{frontmatter}



\dochead{}

\title{Measurement of Top Quark Properties in Single Top-Quark Production at CMS}


\author{Efe Yazgan for the CMS Collaboration}

\address{Ghent University, CERN PH-UCM Bat 42 2-029 C28810, CH 1211 Geneva 23, Switzerland}

\begin{abstract}
Single top-quark t-channel production is exploited for studies of top quark properties. 
The analyses include the measurement of the CKM matrix element, $|V_{tb}|$, search for anomalous couplings of the top quark using a Bayesian neural network  analysis, measurement of single top-quark polarization which directly confirms the V-A nature of the $tWb$ production vertex, and the measurement of W-helicity fractions in the phase space sampled by a selection optimized for t-channel single top-quark production, orthogonal to the $t\overline{t}$ final states used in traditional measurements of these properties. All measurements are found to be consistent with the standard model predictions. 
\end{abstract}


\end{frontmatter}


\section{Introduction}
\label{sec:intro}
The top quark is the most massive particle known to date. The top quark decays via the weak interaction and due to its high mass, it has a very short lifetime which is smaller than the hadronization time-scale, $1/\Lambda_{QCD}$. Therefore, top quark properties can be measured before being obscured by QCD effects. 
Single top-quarks are produced through the electroweak interaction. At the leading order, W boson virtuality is used to classify the single top-quark production in $s$-, $t$-, and $Wt$-channels. Single top-quark production is first observed in  2009 by both Tevatron experiments in the $s+t$ channel using multi-variate techniques \cite{ref:tev1,ref:tev2}. In the subsequent analyses by  Tevatron and LHC experiments, all production modes are established \cite{ref:tchan_D0,ref:twchan_CMS,ref:schan_D0CDF}. Single top-quark measurements provide tests of electroweak interactions, and single top-quark production is sensitive to up and down quark Parton Distribution Functions (PDFs). All three production modes are sensitive to $tWb$-vertex and hence, new physics. For example, $tW$- or $s$-channel production is useful in $W'$ and charged Higgs searches and t-channel production can be used to look for flavor changing neutral currents (FCNCs). In addition, single top-quark is background to Higgs boson and new physics searches. 

\begin{figure}[h!bt] 
\centerline{\includegraphics[width=7.5cm]{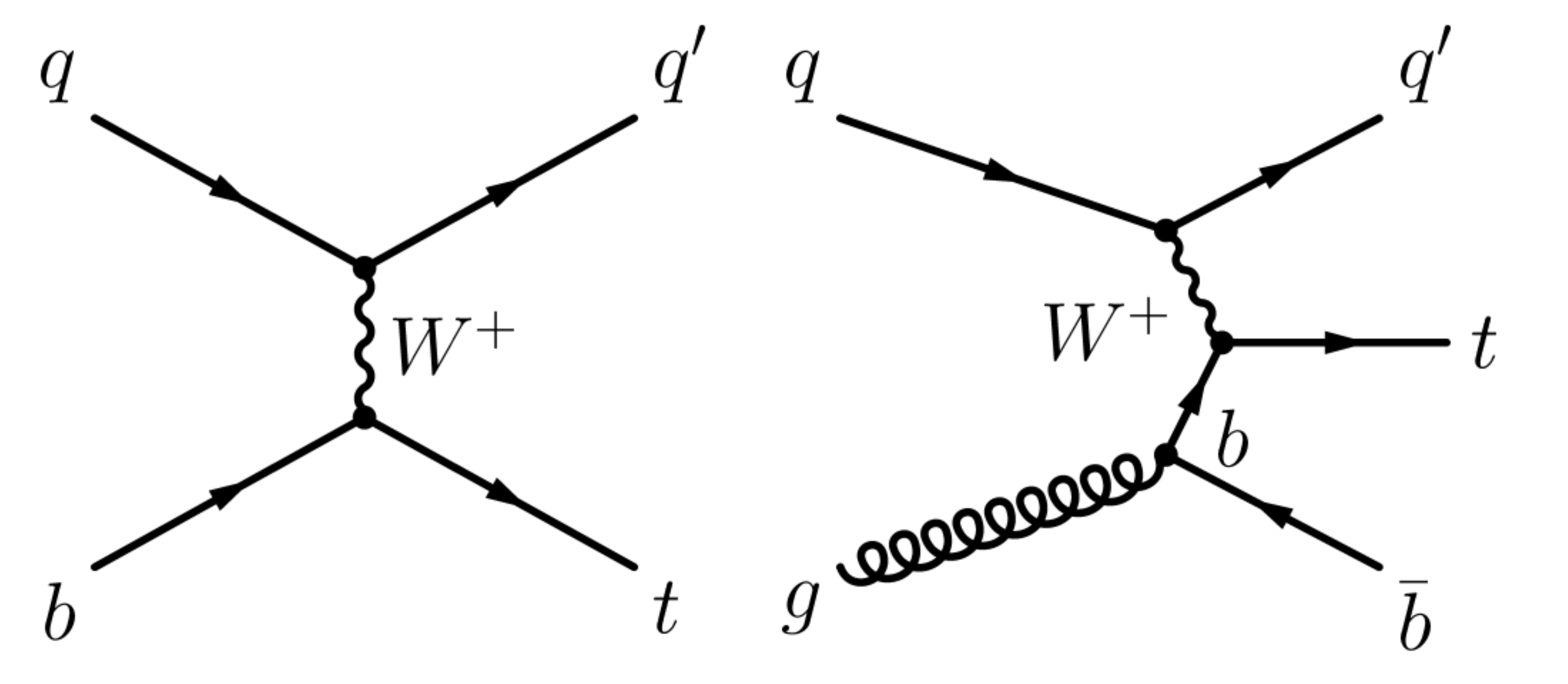}}
\caption{Leading order Feynman diagrams for t-channel top quark production.}
\label{fig:feynman} 	
\end{figure}

The dominant single top-quark process at the LHC (and at the Tevatron) is the t-channel production, and therefore top quark properties measurements using single top-quark production are made in this channel. The Feynman diagram for this process is displayed in Figure \ref{fig:feynman}.
 The t-channel production is characterized by the existence of one isolated lepton, one light and relatively forward jet, one central b-jet and missing transverse energy ($\cancel{\it{E}}_{T}$). The main backgrounds are W+jets, $t\overline{t}$ and QCD multi-jets. To test the standard model (SM) couplings in single top-quark t-channel, an effective field theory approach is used. The most general, lowest dimension, CP-conserving Lagrangian \cite{ref:lag1,ref:lag2} describing the $tWb$-vertex  can be written as
\begin{multline}
\mathcal{L} = -\frac{g}{\sqrt{2}}\overline{b}\gamma^\mu(f_V^LP_L+f_V^RP_R)tW_\mu^- \\ 
-\frac{g}{\sqrt{2}}\overline{b}\frac{i\sigma^{\mu\nu}{\partial}_\nu W_\mu^-}{M_W}(f_T^LP_L+f_T^RP_R)t+h.c.
\label{eq:1}
\end{multline}
where $f_V^L$ and  $f_V^R$ represent the left and right vector operators and $f_T^L$ and  $f_T^R$ are the left and right tensor operators, $P_{L,R}=(1\mp\gamma_5)/2$, $\sigma^{\mu\nu}=i(\gamma_\mu\gamma_\nu-\gamma_\nu\gamma_\mu)/2$. In the SM, at the tree level, $f_V^L=V_{tb}$, and $f_V^R=f_T^L=f_V^R=0$. Section \ref{sec:vtb} presents the measurement of $|V_{tb}|$ assuming $f_V^R=f_T^L=f_V^R=0$. Section \ref{sec:anom} presents the searches for non-zero $|f_V^R|$, $|f_T^L|$, $|f_V^R|$ or $|f_V^L|\neq1$. These measurements are sensitive to the $tWb$-vertex both in production and in decay. Sections \ref{sec:polarization} and \ref{sec:helicity} summarize the top quark and W boson polarization  measurements that test the V-A coupling and which are sensitive to $tWb$-vertex in production and in decay, respectively. 

\section{Measurements of $|V_{tb}|$}
\label{sec:vtb}
\begin{figure}[h!bt] 
\centerline{\includegraphics[width=7.5cm]{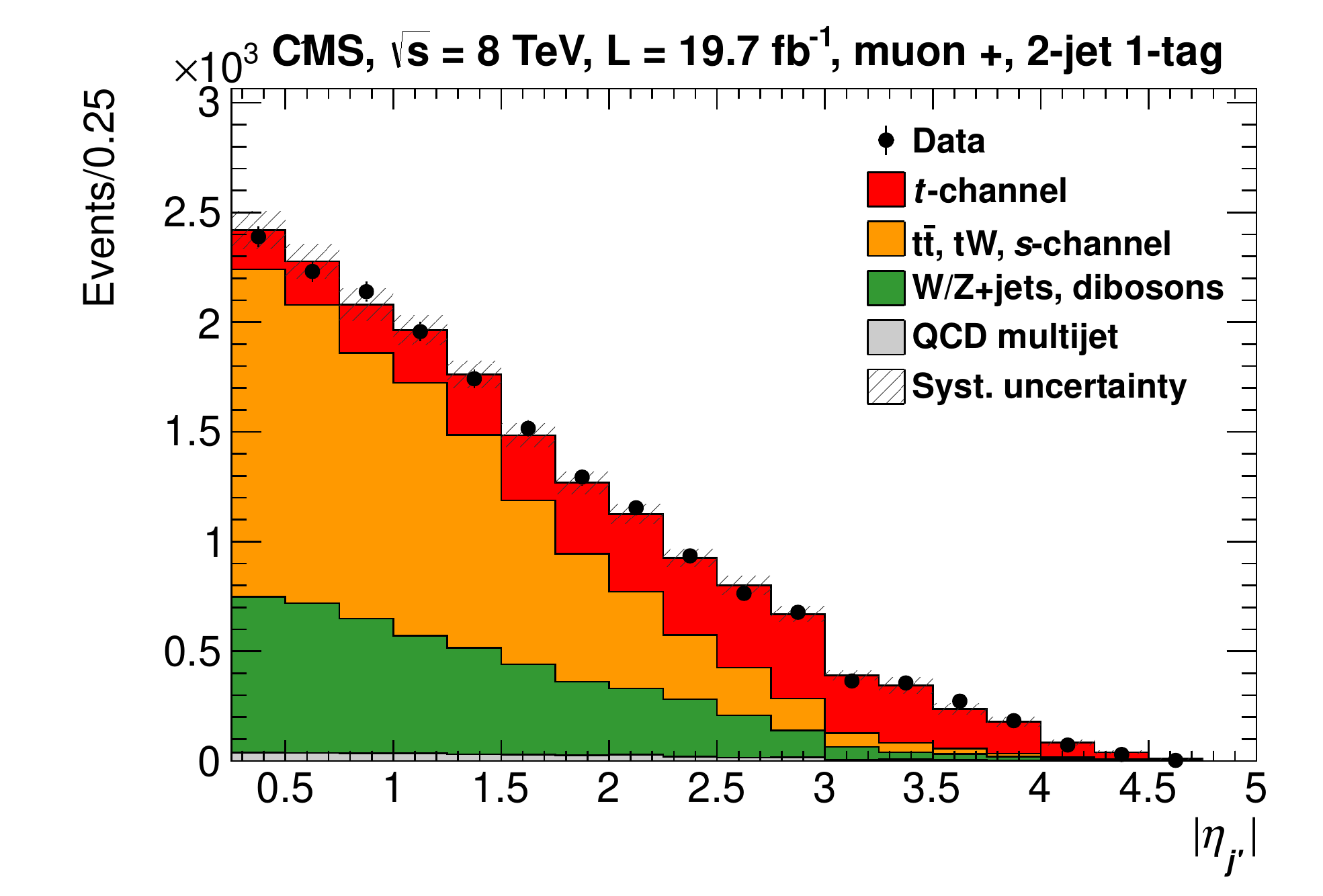}}
\caption{Fitted distribution of $|\eta_{j'}|$ for the $\mu^+$ channel.}
\label{fig:etarecoiljet} 	
\end{figure}

\begin{figure}[h!bt] 
\centerline{\includegraphics[width=7.5cm]{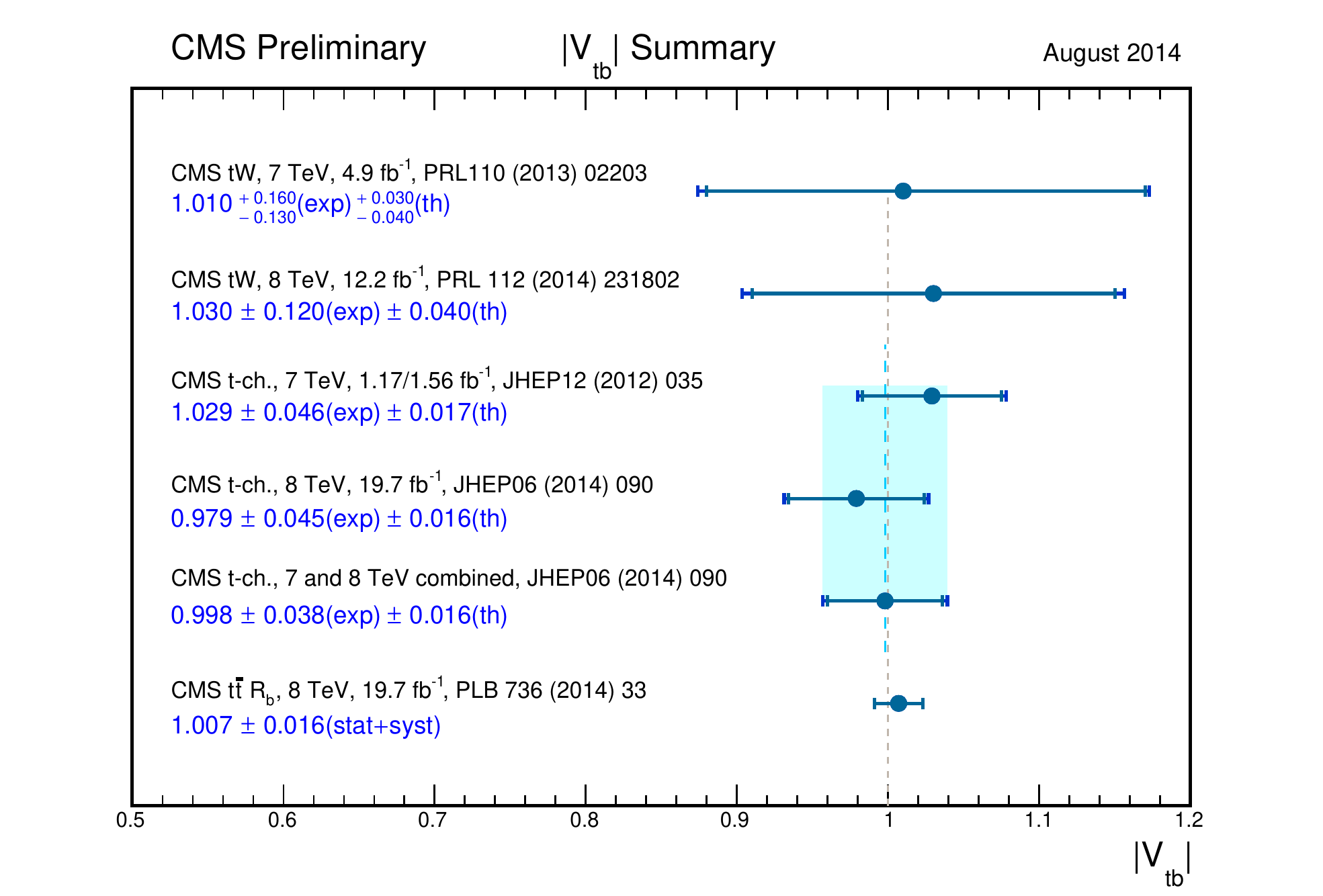}}
\caption{Summary of $|V_{tb}|$ measurements made by CMS.}
\label{fig:cms_vtb} 	
\end{figure}

Some new physics models predict $f_V^L\neq1$ but only an insignificant modification in the branching ratio. Therefore, using the fact that $|V_{tb}|$ is much larger than $|V_{td}|$ and $|V_{ts}|$, an anomalous coupling at the $tWb$-vertex  can be parametrized by $f_V^L$ and can be related to the measured ($\sigma_{t-ch.}$) and the theoretical ($\sigma_{t-ch.}^{theo.}$) t-channel cross sections using the following equation
\begin{equation}
|f_V^LV_{tb}|=\sqrt{\frac{\sigma_{t-ch.}}{\sigma_{t-ch.}^{theo.}}}
\label{eq:fvlvtb}
\end{equation}
 The t-channel cross-section is determined by a maximum likelihood fit to the absolute value of the pseudo-rapidity distribution of the recoiling light jet ($|\eta_j'|$) \cite{ref:vtb} for which the signal is more dominant in the forward region. The  $\eta_j'$ distribution is displayed in Figure \ref{fig:etarecoiljet}. Using the measured cross-section and the estimated cross section at NNLO+NNLL precision \cite{ref:sigmat-ch_theo1,ref:sigmat-ch_theo2}, $|f_V^LV_{tb}|$ is found to be 0.979$\pm$0.045(exp.)$\pm$0.016(theo.) using CMS \cite{ref:CMS} data at $\sqrt{s}$= 7 TeV and  0.998$\pm$0.038(exp.)$\pm$0.016(theo.) using the combined 7 and 8 TeV data. For the combined measurement, the total uncertainty is 4.1\% and the measurement is limited by the statistical uncertainty. Assuming $f_V^L=1$, and $|V_{tb}|\leq$1, a lower limit of $|V_{tb}|$ is obtained as $0.92~@~95\%$ C.L. \cite{ref:vtb}. Summary of $|V_{tb}|$ measurements made by CMS including the measurement in the $t\overline{t}$ production \cite{ref:rmeas} is shown in Figure \ref{fig:cms_vtb}. 

\section{Anomalous Couplings in the t-channel Single-Top Production}
\label{sec:anom}
\begin{figure}[h!bt] 
\centerline{\includegraphics[width=7.5cm]{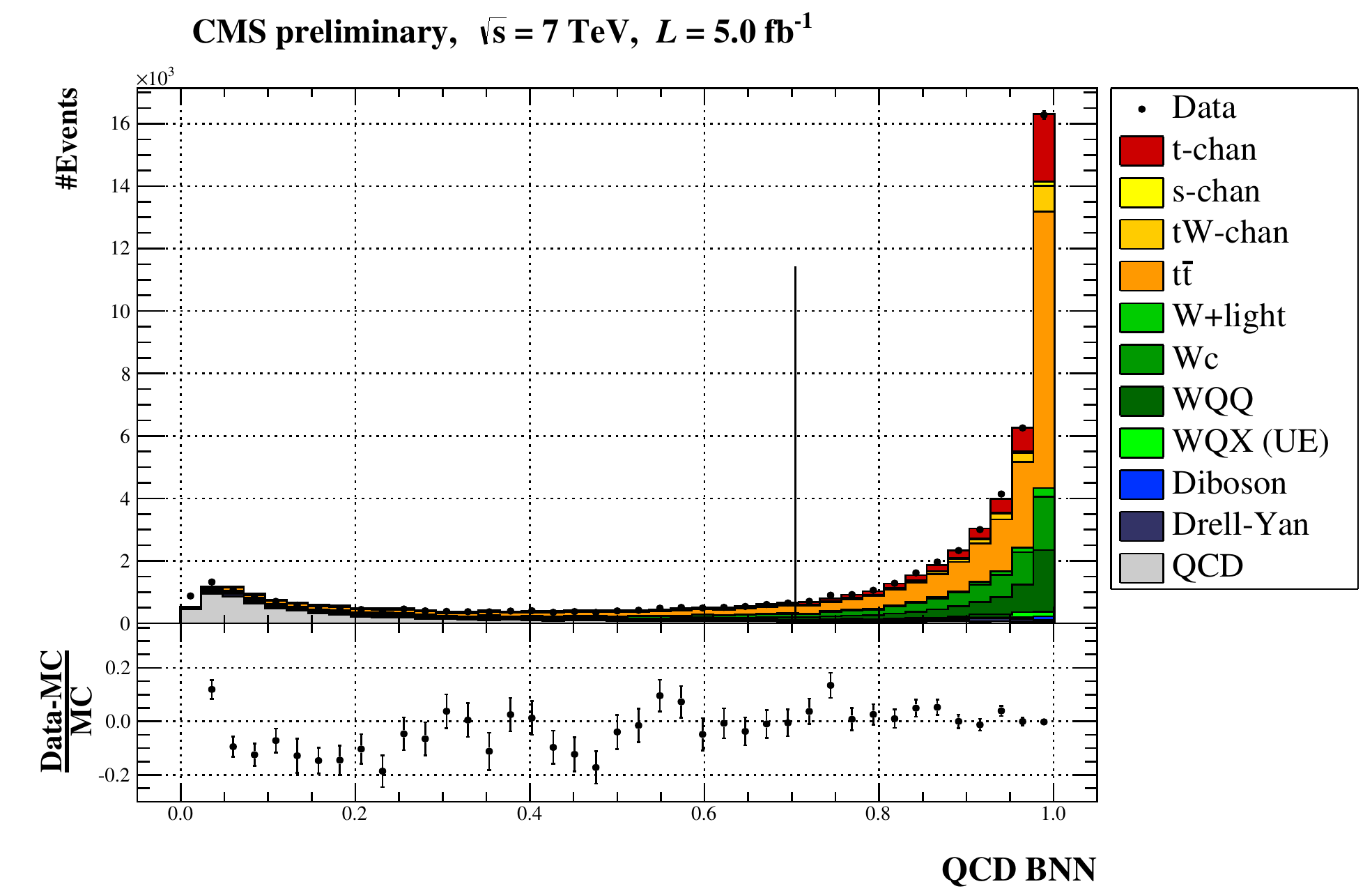}}
\caption{BNN discriminator for the simulation and the data to reject the QCD background shown with only the statistical error bars. The vertical line on the plot indicates the cut value used in the analysis. The error bars on the data points represent the statistical errors.}
\label{fig:QCD_BNN} 	
\end{figure}

\begin{figure}[h!bt] 
\centerline{\includegraphics[width=7.5cm]{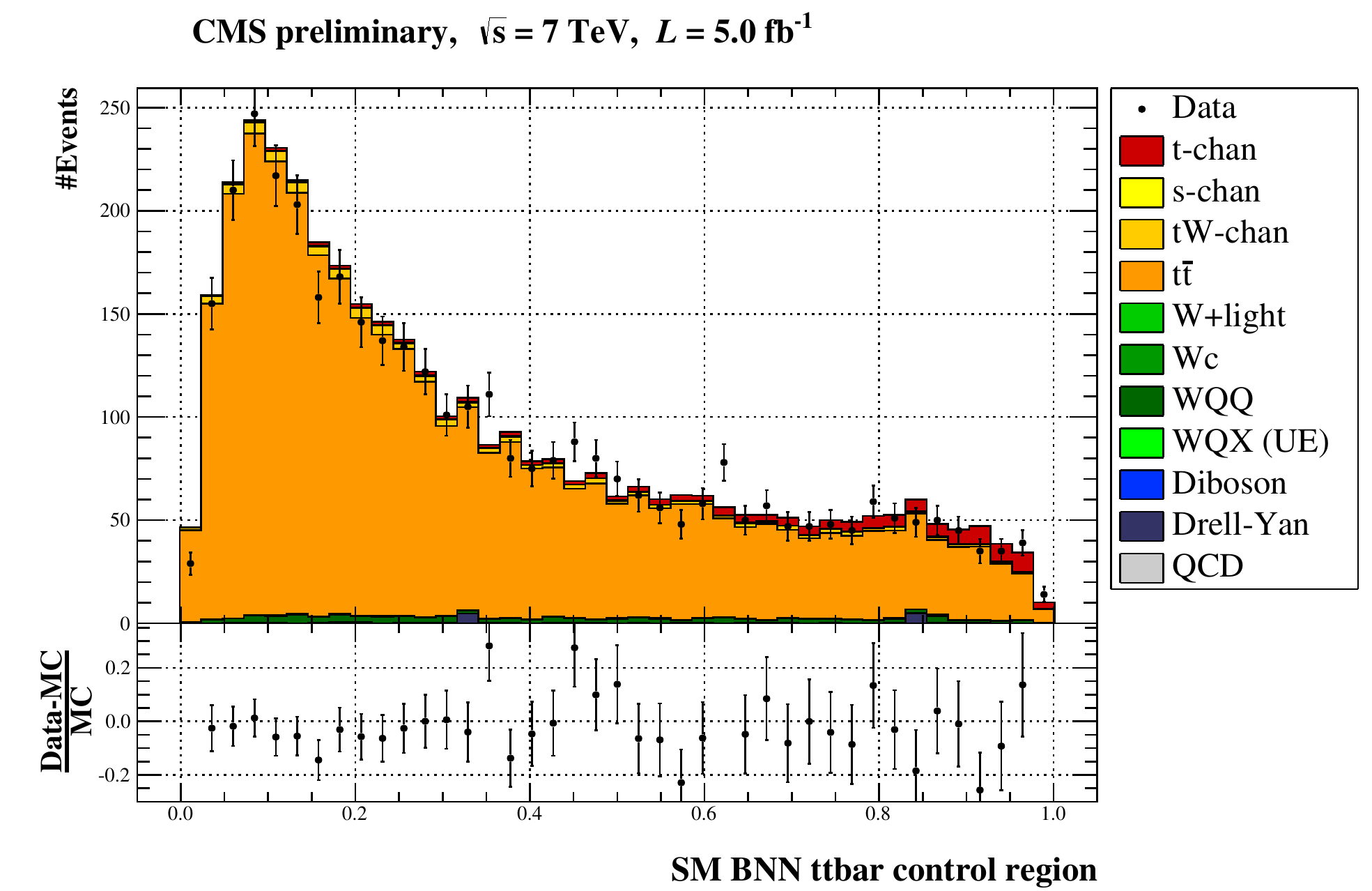}}
\caption{SM BNN discriminator in data and simulation in the $t\overline{t}$ control region. The error bars on the data points represent the statistical errors.}
\label{fig:ttbar} 	
\end{figure}

\begin{figure}[h!bt] 
\centerline{\includegraphics[width=7.5cm]{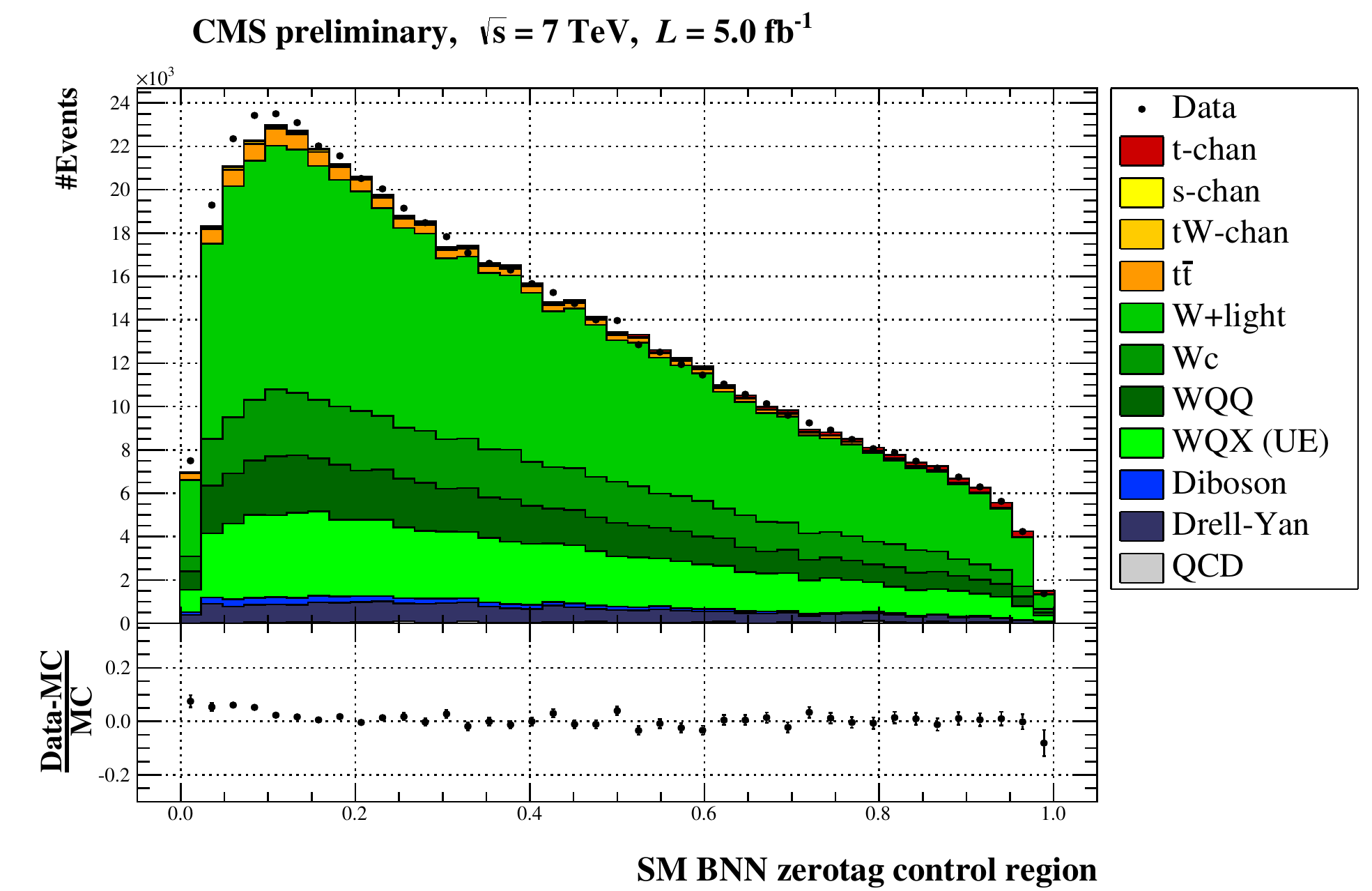}}
\caption{SM BNN discriminator in data and simulation in the W+jets control region. The error bars on the data points represent the statistical errors.}
\label{fig:wjets} 	
\end{figure}

\begin{figure}[h!bt] 
\centerline{\includegraphics[width=7.5cm]{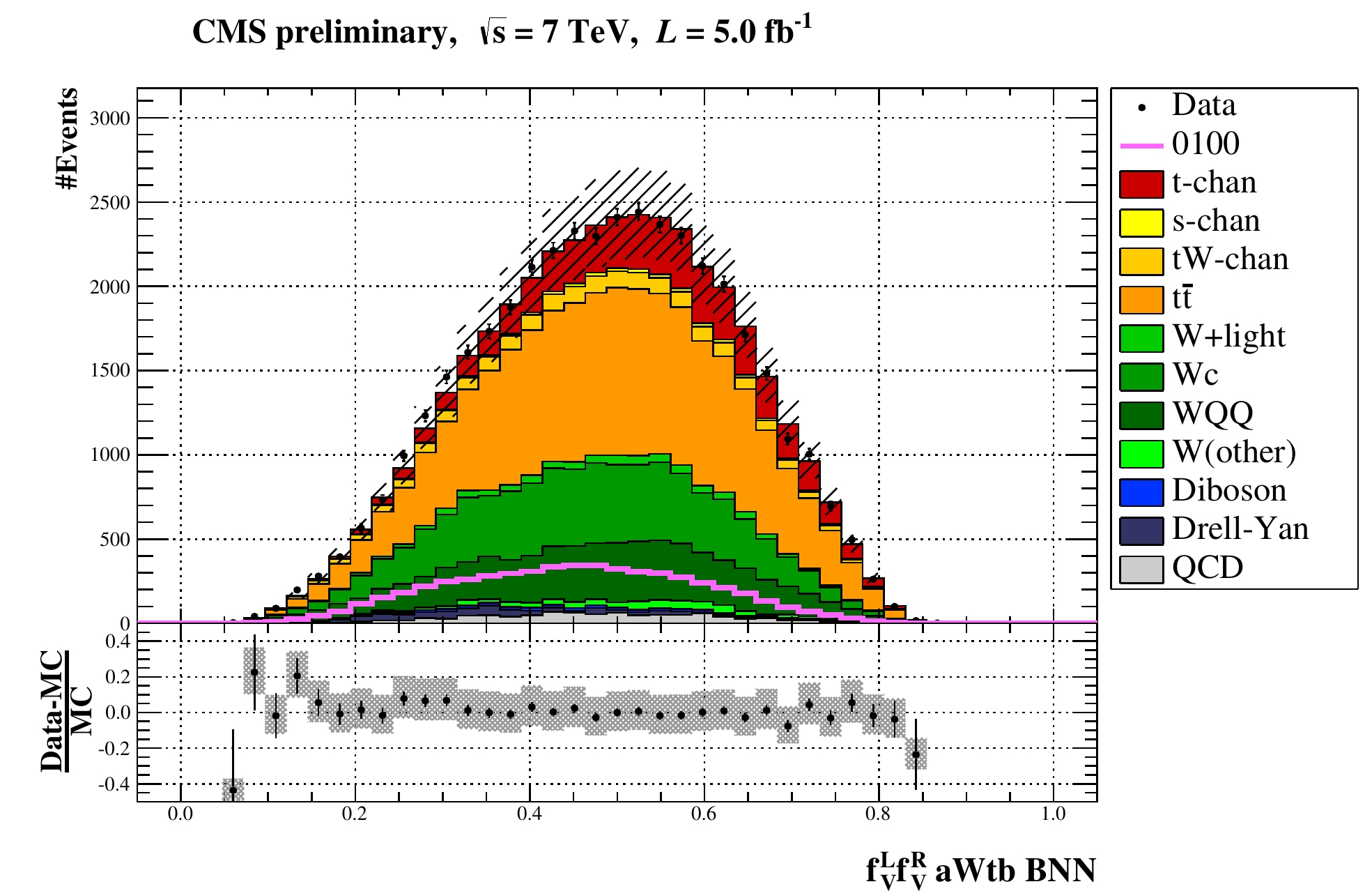}}
\caption{BNN discriminant for the $f_V^L,f_V^R$ scenario. The hashed band represents the systematic uncertainty. The error bars on the data points represent the statistical errors.}
\label{fig:fvlfvr} 	
\end{figure}

\begin{figure}[h!bt] 
\centerline{\includegraphics[width=7.5cm]{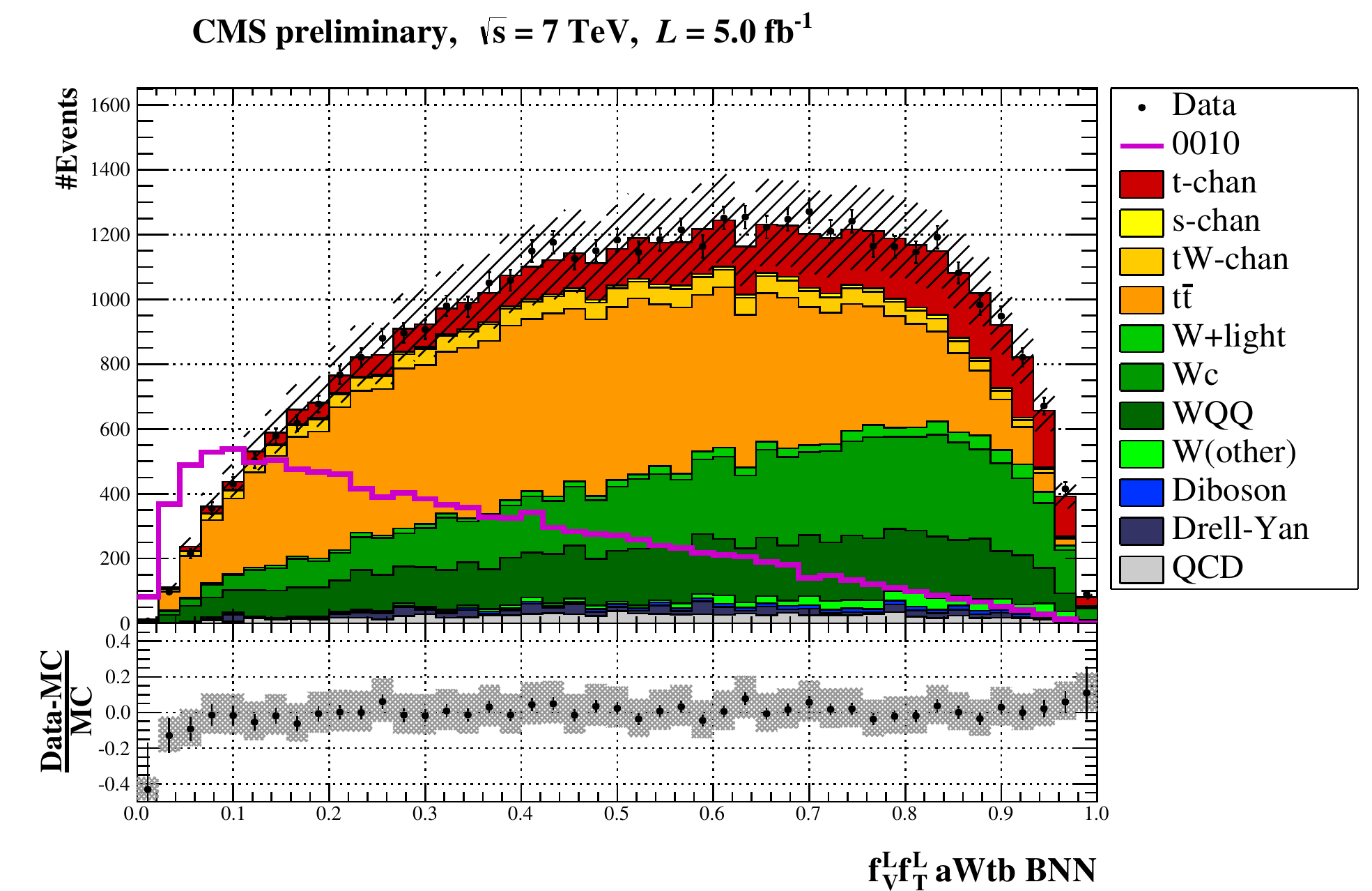}}
\caption{BNN discriminant for the $f_V^L,f_T^L$ scenario. The hashed band represents the systematic uncertainty.}
\label{fig:fvlftl} 	
\end{figure}

Analyzing data corresponding to an integrated luminosity of 5 fb$^{-1}$ in the muon+jets channel, a search for anomalous couplings is made in the t-channel \cite{ref:anom_cms}.  Separate Bayesian Neural Networks (BNN) using up to 25 variables are used to suppress the QCD multi-jet background, discriminate signal and backgrounds, and finally to search for anomalous $tWb$-couplings. The BNN to discriminate SM signal from the QCD multi-jet background is displayed in Figure \ref{fig:QCD_BNN}. To discriminate the SM signal from the $t\overline{t}$ background, a BNN control region defined by 4 jets with 1 b-tagged jets (Figure \ref{fig:ttbar}) and from the $W+jets$ background, a BNN control region defined by zero b-tagged jets (Figure \ref{fig:wjets}) are used. To search for anomalous couplings two of the four couplings in Equation \ref{eq:1} are analyzed simultaneously: $(f_V^L,f_V^R)$, $(f_V^L,f_T^L)$, and $(f_V^L,f_T^R)$ fixing the other two couplings to zero in each case.  Other BNNs are trained to  distinguish SM $f_V^L$  and anomalous $tWb$-couplings $f_V^R$ (Figure \ref{fig:fvlfvr}) and $f_T^L$ (Figure \ref{fig:fvlftl}). The observed event yields are found to be consistent with the SM predictions and the following exclusion limit pairs are derived at 95\% C.L.: $|f_V^L|>0.90~(0.88),~|f_V^R|<0.34~(0.39)$ and $|f_V^L|>0.92~(0.88),~|f_T^L|<0.09~(0.16)$. The numbers in the parentheses represent the expected limits. 

\section{Top Quark Polarization}
\label{sec:polarization}
\begin{figure}[h!bt] 
\centerline{\includegraphics[width=7.5cm]{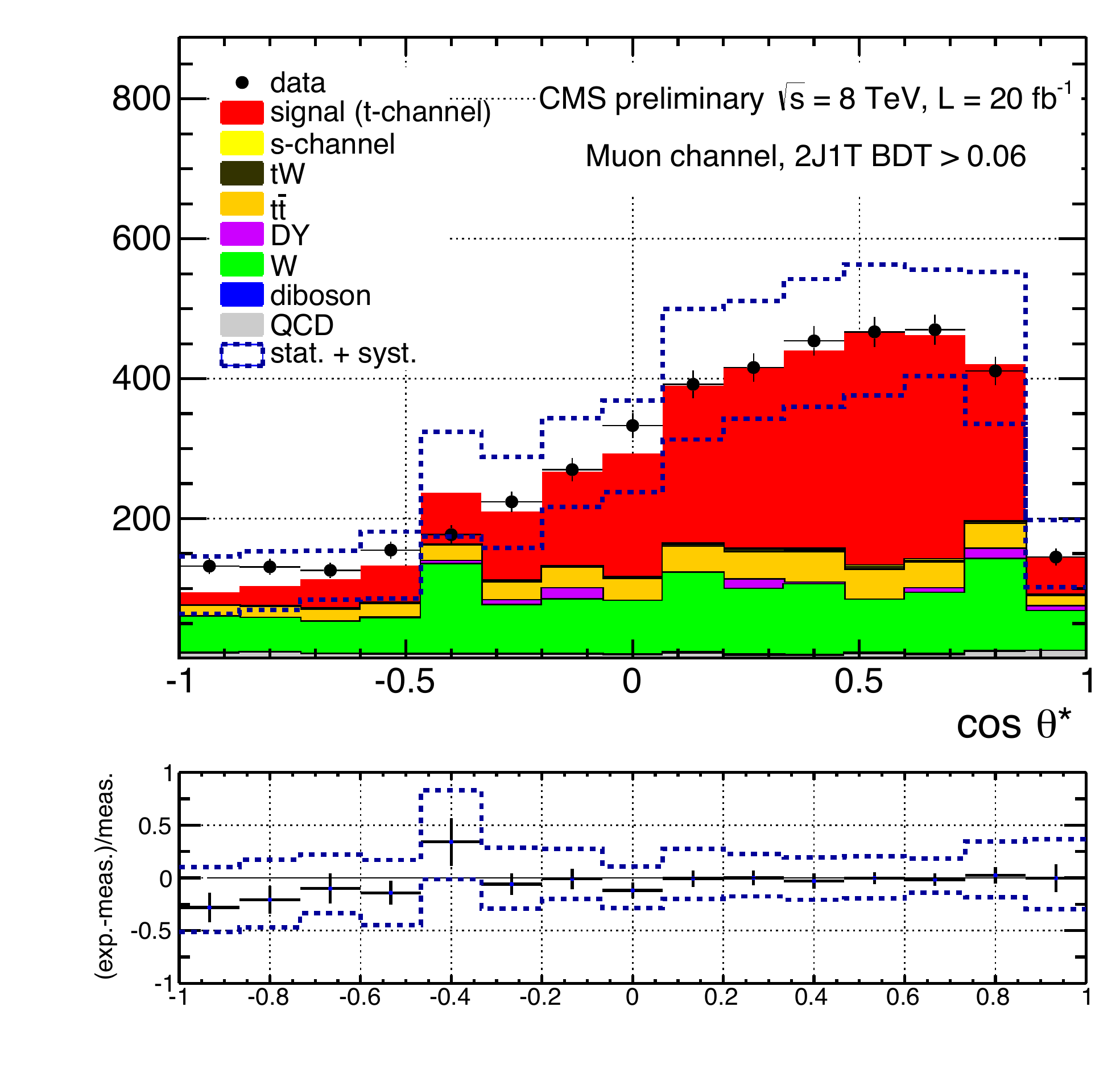}}
\caption{The post-fit $\cos\theta_\ell^*$ distribution in the muon channel for events that passed the BDT discriminant cut. The dashed lines indicate the combined statistical and systematic uncertainty on the MC.}
\label{fig:pol1} 	
\end{figure}

\begin{figure}[h!bt] 
\centerline{\includegraphics[width=7.5cm]{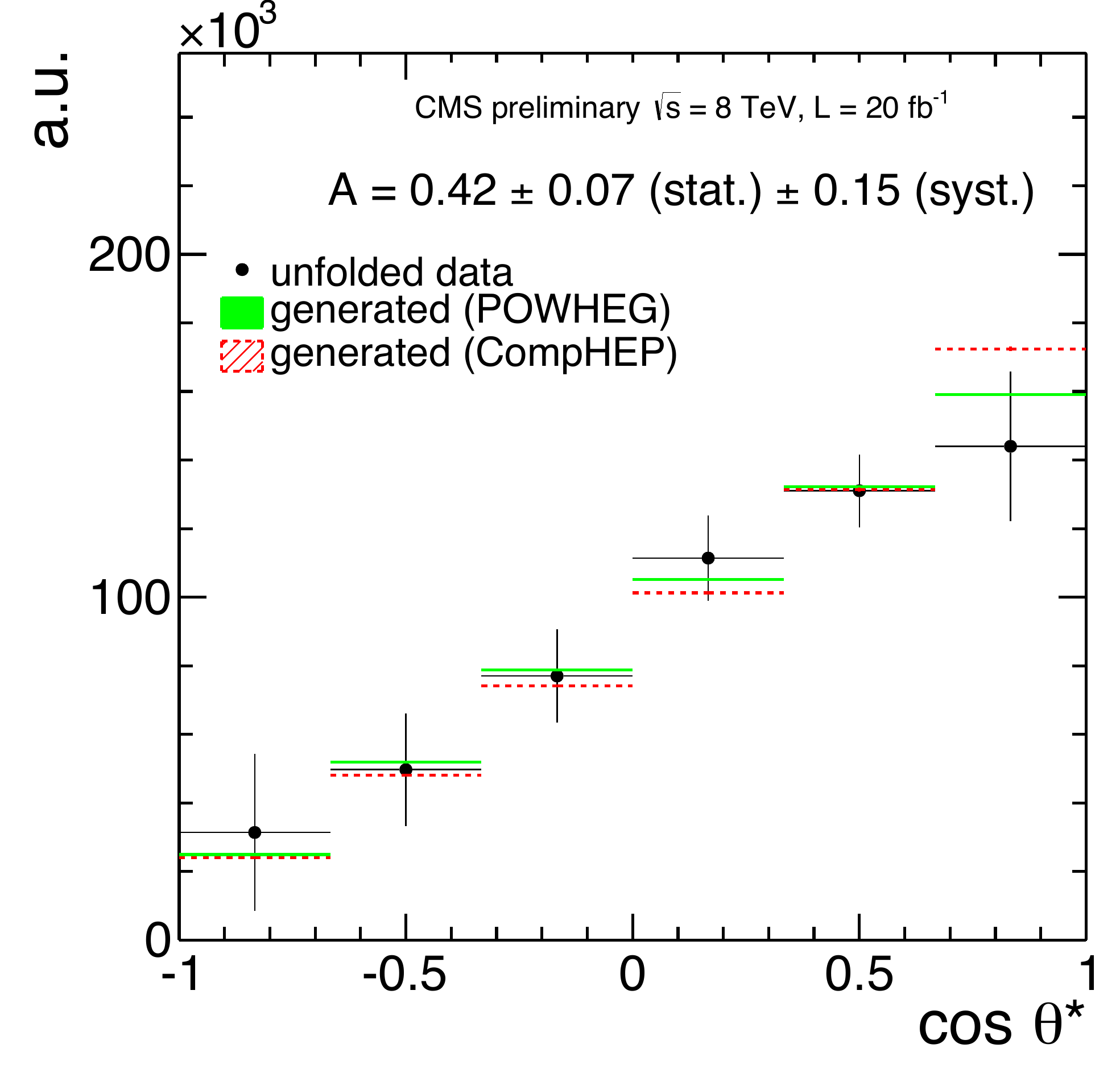}}
\caption{The unfolded $\cos\theta_\ell^*$ distribution in the muon channel shown with the predictions from POWHEG and CompHEP. The angle $\theta_\ell^*$ is defined to be the angle between the charged lepton and the light-quark jet. The error bars indicate the statistical uncertainties.}
\label{fig:pol2} 	
\end{figure}

Single top-quarks are produced mostly with left-handed polarization in the SM due to the V-A coupling. New particles or interactions could possibly modify the top quark polarization to be less than $\sim100$\%. At the parton-level, the angular distribution ($\theta_X$) of a decay product X is given by \cite{ref:mahlonparke, ref:jezabek1994}

\begin{equation}
\frac{d\Gamma}{d\cos\theta_X}=\frac{\Gamma}{2}(1+P_t\alpha_X\cos\theta_X)\equiv\Gamma(\frac{1}{2}+A_X\cos\theta_X)
\end{equation}

where X is $W,\ell,\nu$ or $b$, $\Gamma$ is the partial decay with of X, $P_t$ is the single top-quark polarization, and $\alpha_X$ is the the spin-analyzing power. In this equation $A_X$ is the forward-backward asymmetry of the decay product X.
CMS made a measurement of top quark polarization in the t-channel single top-quark events \cite{ref:top_pol}. In this analysis, the polarization is extracted using the slope of the $\cos\theta_\ell^*$ distribution unfolded to the parton-level. The angle $\theta_\ell^*$ is defined between the charged lepton and the not-b-tagged jet (i.e. the light-quark jet) in the top quark rest frame. This choice was motivated by the fact that, at the production vertex, the final state light-quark tends to have a direction parallel to the spin of the top quark \cite{ref:mahlonparke}. In the measurement, a boosted decision tree (BDT) that is fit to the data is used to extract the signal and background. The reconstructed and unfolded $\cos\theta_\ell*$ distributions in the muon channel are displayed in Figures \ref{fig:pol1} and \ref{fig:pol2}, respectively. The unfolded data in Figure \ref{fig:pol2} is compared to the predictions from POWHEG \cite{ref:powheg} and CompHEP \cite{ref:comphep1,ref:comphep2}. From the unfolded $\cos\theta^*$ distributions, the forward-backward asymmetry $A_\ell$ of the leptons in the top quark rest frame is calculated using the equation
\begin{equation}
A_\ell=\frac{N(\cos\theta^*>0)-N(\cos\theta^*<0)}{N(\cos\theta^*>0)+N(\cos\theta^*<0)}
\end{equation}
The asymmetry in the muon and the electron channels are
\begin{eqnarray}
A_\mu=0.42\pm0.07~(stat)\pm0.15~(sys),\\
A_e=0.31\pm0.11~(stat)\pm0.23~(sys).
\end{eqnarray}
The electron and muon channel results are combined using the BLUE method \cite{ref:blue}. The combination yields 
\begin{equation} 
A_\ell=\frac{\alpha_\ell P_t}{2}=0.41\pm0.06~(stat)\pm0.16~(sys). 
\end{equation}
Assuming that the spin analyzing power of the charge lepton ($\alpha_\ell$) is unity, the measured degree of polarization is $P_t=0.82\pm0.12~(stat)\pm0.32~(sys)$ which is in agreement with the SM V-A coupling.
The measurement is dominated by the uncertainties in the jet energy scale, factorization scale, top quark mass and the background estimation.  

\section{W Helicity Fractions in Single Top-Quark Topologies}
\label{sec:helicity}
\begin{figure}[h!bt] 
\centerline{\includegraphics[width=7.5cm]{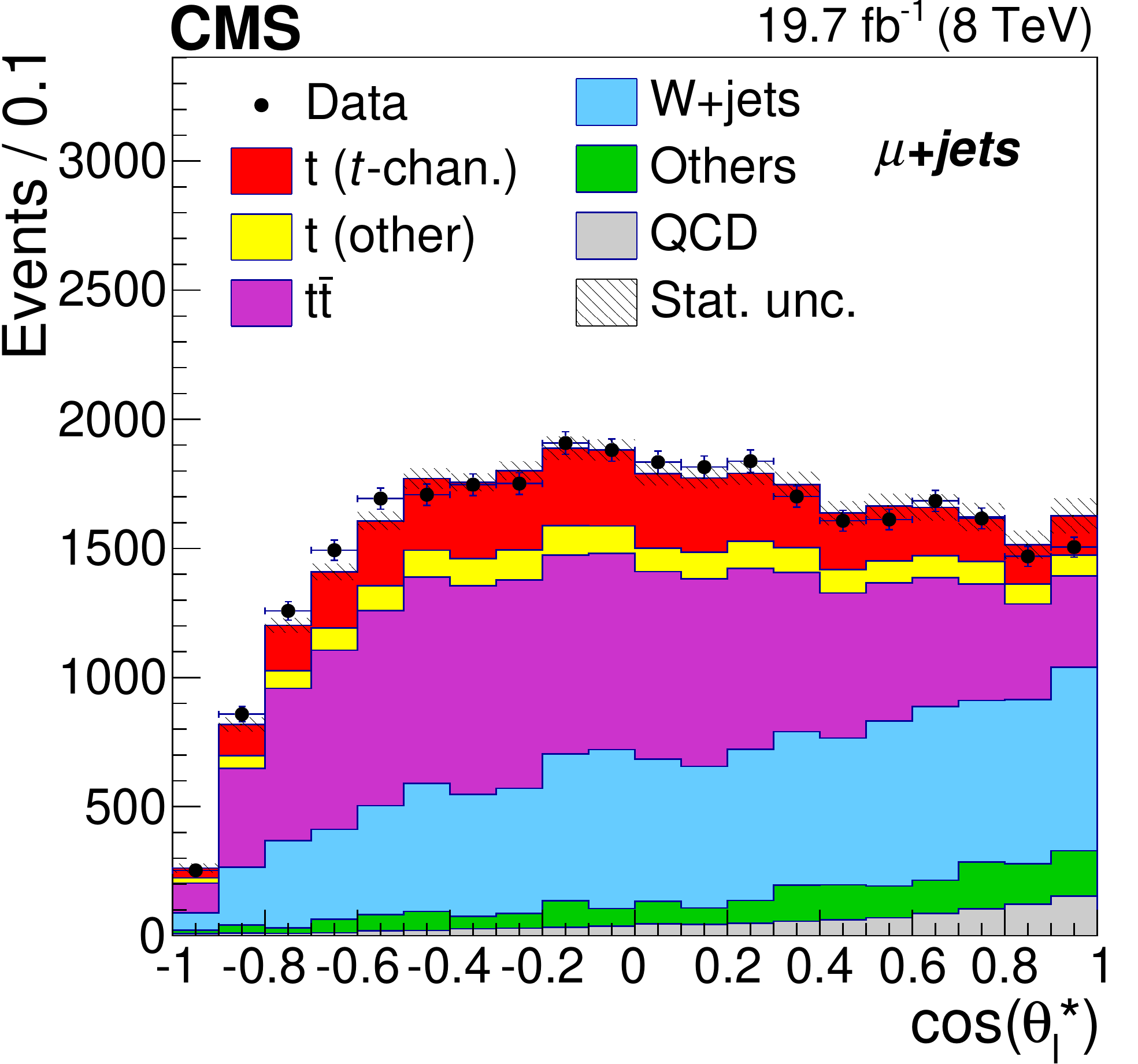}}
\caption{ The reconstructed $\cos\theta_\ell^*$ for data and simulation at $\sqrt{s}=$ 8 TeV in the muon channel. The angle $\theta_\ell^*$ is defined between the momentum of the down-type fermion in the W rest-frame and the W momentum in the top quark rest frame. }
\label{fig:whel_7} 	
\end{figure}

\begin{figure}[h!bt] 
\centerline{\includegraphics[width=7.5cm]{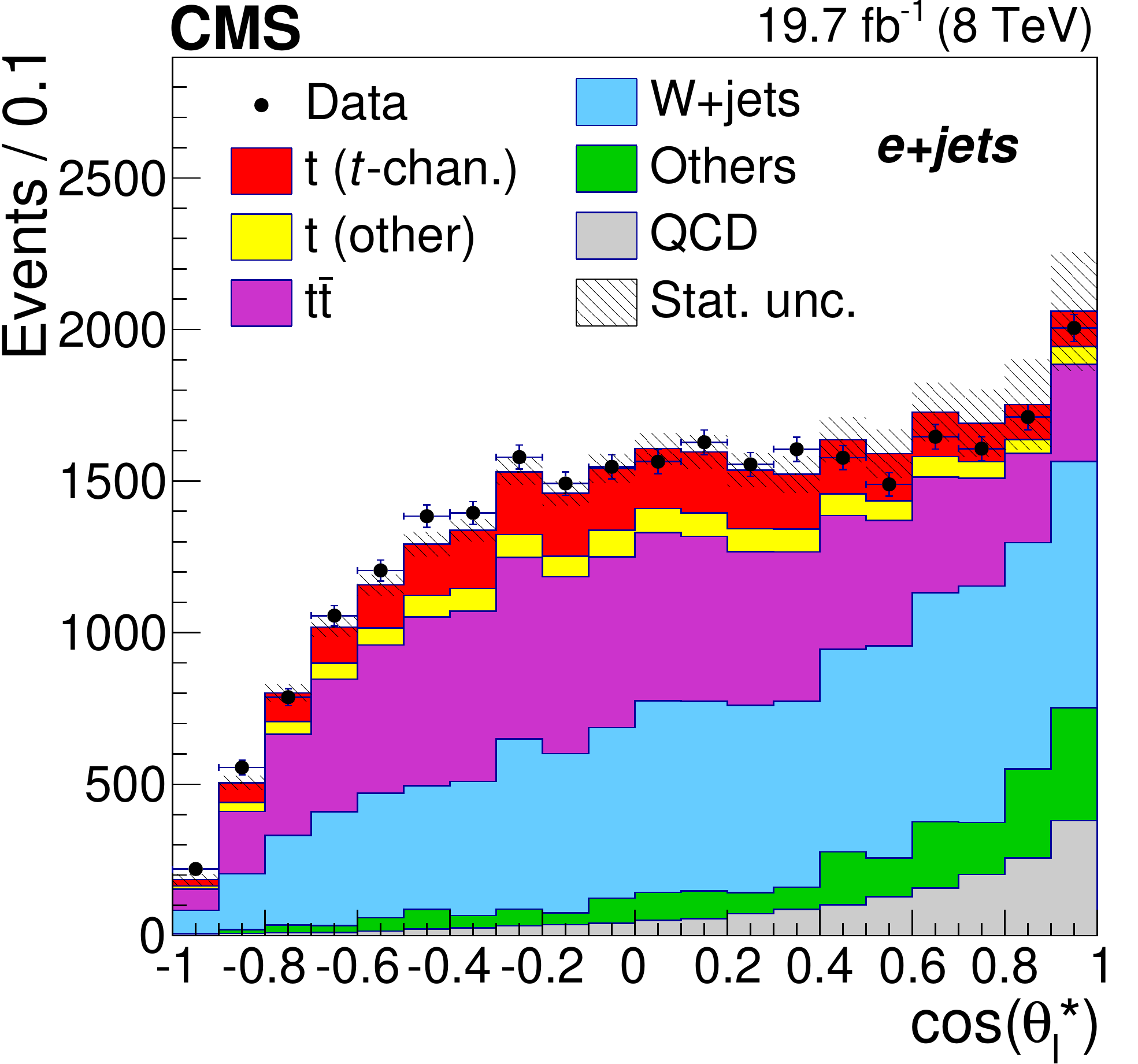}}
\caption{ The reconstructed $\cos\theta_\ell^*$ for data and simulation at $\sqrt{s}=$ 8 TeV in the electron channel. The angle $\theta_\ell^*$ is defined between the momentum of the down-type fermion in the W rest-frame and the W momentum in the top quark rest frame. }
\label{fig:whel_8} 	
\end{figure}

\begin{figure}[h!] 
\centerline{\includegraphics[width=7.cm]{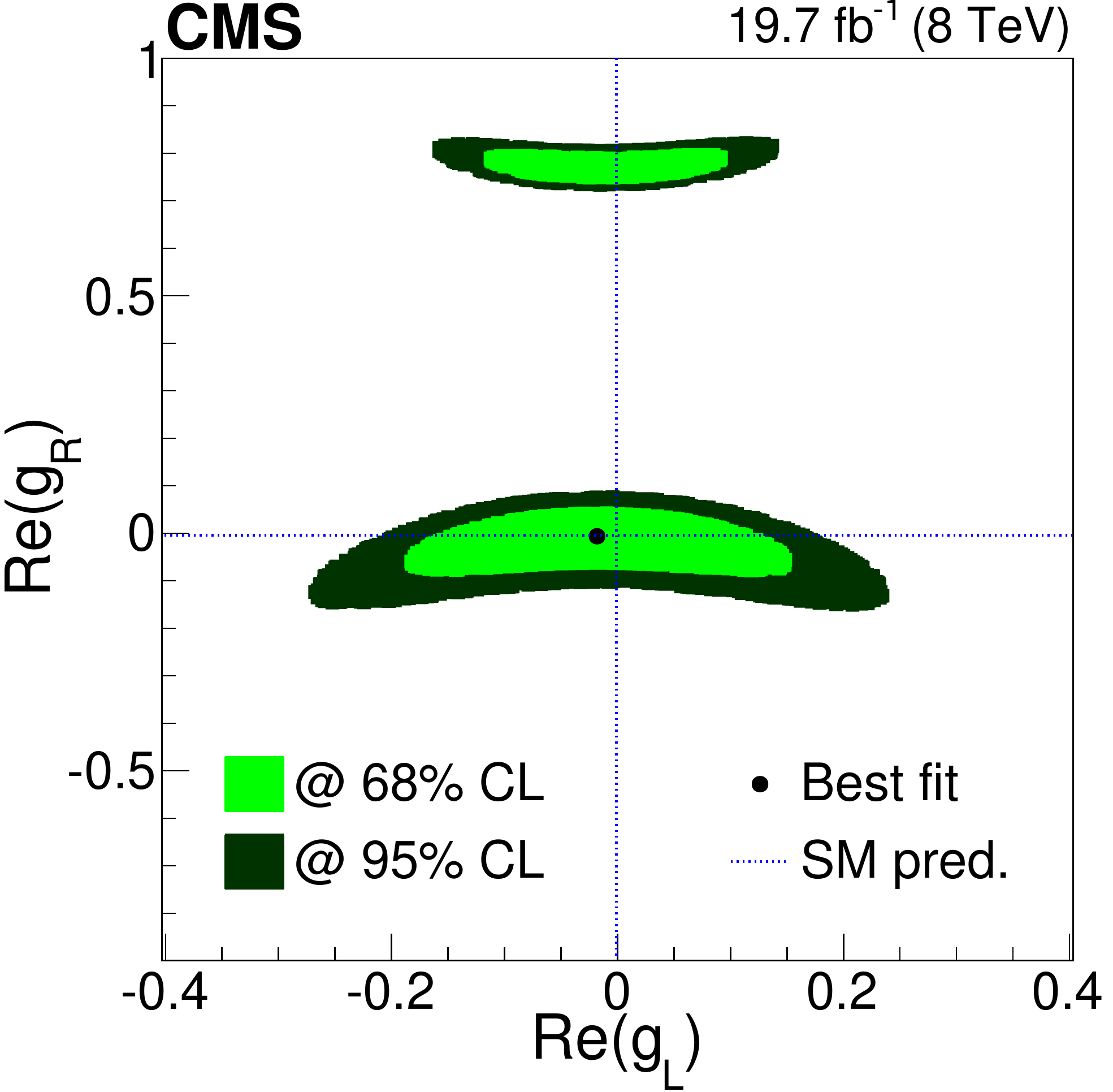}}
\caption{Exclusion limits with 68\% and 95\% confidence limit on real left and right anomalous tensor couplings using the combined W-helicity measurements in single top-quark event topology.}
\label{fig:glgr} 	
\end{figure}

In the top quark decay, the W boson can be produced in three possible helicity states: left-handed ($F_L$), longitudinal ($F_0$), and right-handed ($F_R$) with $F_L+F_0+F_R=1$.
The V-A nature of the top quark decay implies that (a massless) b-quark is produced left-handed, and therefore 
W bosons can not be produced in a right-handed state because of the angular momentum conservation. 
At NNLO, with b-quark mass, $m_b=4.8$ GeV and top quark mass, $m_t=172.8~(1.3)$ GeV, the helicity fractions are $F_0=0.687~(5)$, $F_L=0.311~(5)$, $F_R=0.0017~(1)$ \cite{ref:czarnecti2010}. The W helicities can be measured utilizing the angle ($\theta^*$) between the momentum of the down-type fermion in the W rest-frame and the W momentum in the top quark rest-frame. The $\theta^*$ distribution is given by
\begin{multline}
\frac{1}{N}\frac{N}{d\cos\theta^*}=\frac{3}{8}F_L(1-\cos\theta^*)^2
+\frac{3}{8}F_R(1+\cos\theta^*)^2\\+\frac{3}{4}F_0\sin^2\theta^*~~~~~~~~~~~~~~~~~~~~~~~~~~~~~~~~~~
\end{multline}
By measuring $\cos\theta^*$, CMS determined the W-helicity fractions using single top-quark events in the $\mu$+jets and $e$+jets final states \cite{ref:whel}. The W boson helicities are obtained from likelihoods with re-weighted signals. The re-weighting includes all processes involving the top quark (i.e. t-, s-, tW- channels, and $t\overline{t}$ lepton+jets and dilepton final states). The helicity fractions and the W+jets background contribution are extracted simultaneously. The $\cos\theta^*$ distributions measured using data taken at 8 TeV in muon+jets and electron+jets channels are shown in Figures \ref{fig:whel_7} and \ref{fig:whel_8}, respectively. The combined 8 TeV measurement yielded the following helicity fractions: $F_L=0.298\pm0.028~(stat)\pm0.032~(sys)$, $F_0=0.720\pm0.039~(stat)\pm0.037~(sys)$, and $F_R=1-F_0-F_L=-0.018\pm0.019~(stat)\pm0.011~(sys)$. 
These results are consistent with the SM NNLO QCD predictions and the measurements in the $t\overline{t}$ channel \cite{ref:whel_ttbar1,ref:whel_ttbar2,ref:whel_ttbar3}.  
In these measurements, the dominant systematic uncertainty arise from modeling. 
The combined W-helicity measurements are used to determine upper limits on the real parts of the left and right anomalous tensor couplings as displayed in Figure \ref{fig:glgr}. 

\section{Summary and Conclusions}
\label{sec:conc}
Top quark properties measurements from single top-quark data, namely $|V_{tb}|$, anomalous couplings, top quark polarization, and W-boson helicity, are presented. Measurements of top quark properties in single top-quark production at CMS are providing thorough tests of the standard model. All measurements show good agreement with the standard model predictions. 



\nocite{*}
\bibliographystyle{elsarticle-num}
\bibliography{martin}



\end{document}